 \documentclass[smallabstract,smallcaptions]{dccpaper}

\usepackage{epsfig}
\usepackage{amsmath}
\usepackage{amssymb}
\usepackage{color}
\usepackage{url}
\usepackage{cite}
\usepackage{amsmath,amssymb,amsfonts}
\usepackage{algorithmic}
\usepackage{graphicx}
\usepackage{textcomp}
\usepackage{xcolor}
\usepackage{booktabs}
\usepackage{caption}  
\newlength{\figurewidth}
\newlength{\smallfigurewidth}

\begin{document}

\title
{\large
\textbf{From Noise to Latent: Generating Gaussian Latents for INR-Based Image Compression}
}

\author{%
Chaoy Lin$^{\ast}$, Yaojun Wu$^{\ast}$, Yue Li$^{\dag}$, Junru Li$^{\dag}$, Kai Zhang$^{\dag}$, Li Zhang$^{\dag}$ \\[0.5em]
{\small\begin{minipage}{\linewidth}\begin{center}
\begin{tabular}{c}
$^{\ast}$Bytedance Inc., Beijing, China \\
$^{\dag}$Bytedance Inc., San Diego, USA \\
\url{{linchaoyi.cy, zhangkai.video, lizhang.idm}@bytedance.com}
\end{tabular}
\end{center}\end{minipage}}
}

\maketitle
\thispagestyle{empty}

\begin{abstract}
Recent implicit neural representation (INR)-based image compression methods have shown competitive performance by overfitting image-specific latent codes. However, they remain inferior to end-to-end (E2E) compression approaches due to the absence of expressive latent representations. On the other hand, E2E methods rely on transmitting latent codes and requiring complex entropy models, leading to increased decoding complexity. Inspired by the normalization strategy in E2E codecs where latents are transformed into Gaussian noise to demonstrate the removal of spatial redundancy, we explore the inverse direction: generating latents directly from Gaussian noise. In this paper, we propose a novel image compression paradigm that reconstructs image-specific latents from a multi-scale Gaussian noise tensor, deterministically generated using a shared random seed. A Gaussian Parameter Prediction (GPP) module estimates the distribution parameters, enabling one-shot latent generation via reparameterization trick. The predicted latent is then passed through a synthesis network to reconstruct the image. Our method eliminates the need to transmit latent codes while preserving latent-based benefits, achieving competitive rate-distortion performance on Kodak and CLIC dataset. To the best of our knowledge, this is the first work to explore Gaussian latent generation for learned image compression.
\end{abstract}

\section{Introduction}
The rapid advancement of imaging technologies has resulted in an explosion of high-resolution visual data, placing increasing demands on image compression techniques. In recent years, learning-based image compression methods~\cite{c3,coin, balle, minnen, cheng2020} have emerged as a dominant approach, demonstrating superior rate-distortion performance over traditional codecs.

Implicit Neural Representation (INR)-based image compression offers a compression paradigm by representing images as coordinate-conditioned neural networks that directly map spatial locations to pixel values \cite{coin, coinpp, RQAT}. These methods typically do not transmit explicit latent codes and can be optimized per image without large datasets. However, due to the lack of expressive latent representations, they often underperform compared to state-of-the-art end-to-end (E2E) codecs. Recent works such as Cool-Chic~\cite{coolchic_v1,coolchic_v3} and C3~\cite{c3} mitigate this issue by introducing learnable latent tensors that are quantized and included in the bitstream, leading to improved rate-distortion performance. Nevertheless, their performance remains inferior to E2E methods.
\begin{figure}
    \centering
    \includegraphics[width=1\linewidth]{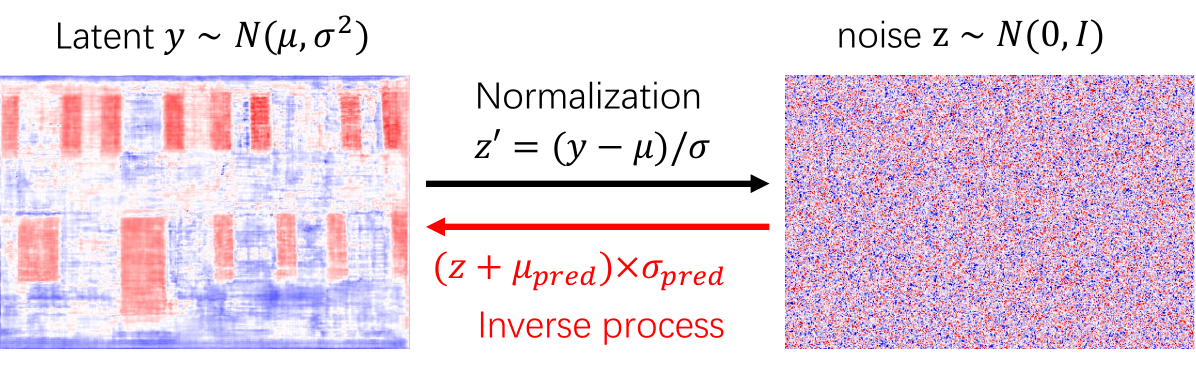}
  \caption{The E2E compression methods transform Gaussian latents \( y \sim \mathcal{N}(\mu, \sigma^2) \) to noise \( z' = (y - \mu)/\sigma \) to demonstrate the effectiveness of spatial redundancy removal. We explore the inverse process: starting from randomly sampled Gaussian noise \( z \sim \mathcal{N}(0, I) \), we estimate the mean \( \mu_{\text{pred}} \) and scale \( \sigma_{\text{pred}} \), and reconstruct the latent as \( y_{\text{pred}} = z \cdot \sigma_{\text{pred}} + \mu_{\text{pred}} \).}
    \label{fig:first_page}
\end{figure}

End-to-end (E2E) codecs~\cite{balle,minnen,cheng2020,elic} typically adopt a variational autoencoder (VAE) framework, where an encoder explicitly extracts a latent representation and a synthesis network reconstructs the image from it. These latents are commonly modeled as Gaussian distributions with learned mean $\mu$ and scale $\sigma$, enabling effective entropy coding. However, achieving compact representations often requires complex auto-regressive entropy models, which introduce significant computational overhead and limit practical deployment.

A key property of E2E codecs is the normalization of latent features into standard Gaussian noise (see Fig.~\ref{fig:first_page}), which illustrates their ability to effectively remove spatial redundancy. Inspired by this, we consider the inverse problem: instead of transforming latents into Gaussian noise, can we start from noise and recover the latent? Specifically, we ask:

\begin{quote}
\textit{Given a randomly sampled Gaussian noise input, is it possible to generate a informative latent representation and reconstruct the image without explicitly storing or transmitting the latent code?}
\end{quote}

In this paper, we explore this question by focusing on the Gaussian latent for INR-based image compression, which modeling the latent as Gaussian distribution. Specifically, we propose a novel INR-based image compression framework that generates image-specific latents from a multi-scale standard Gaussian noise tensor. Both encoder and decoder deterministically construct the same noise tensor using a shared random seed. A lightweight Gaussian Parameter Prediction (GPP) module then predicts the mean and scale of the latent distribution conditioned on the noise. Using the reparameterization trick, we perform one-shot generation of the latent, which is subsequently passed through a synthesis network to reconstruct the image. The entire pipeline is overfitted to a single input image following the INR paradigm. This design removes the need to store or transmit the latent representation, thereby eliminating reliance on computationally intensive auto-regressive entropy models, while preserving high reconstruction quality.

Our contributions are summarized as follows:
\begin{itemize}
    \item We propose a novel compression paradigm for INR-based image compression that reconstructs latents directly from shared noise, eliminating the need to transmit latent codes. To the best of our knowledge, we are the first to explore Gaussian latent generation from fixed noise in the context of INR-based image compression.
    \item We conduct extensive experiments on the Kodak and CLIC dataset, demonstrating that our method achieves superior rate-distortion performance to existing cooridinates-based INR compression methods. Our method offers a lightweight alternative to existing schemes which requires complex auto-regressive process for decoding latents.
\end{itemize}

\section{Motivation and Derivation}
In end-to-end (E2E) image compression, latent features are often modeled as Gaussian latent \( y \sim \mathcal{N}(\mu, \sigma^2) \) as shown in Fig. \ref{fig:first_page}. To facilitate spatial redundancy removal, these latents are normalized into standard Gaussian noise \( z' \sim \mathcal{N}(0, I) \) using the following transformation:
\begin{equation}
    z' = \frac{y - \mu}{\sigma}.
\end{equation}
This normalization enables effective entropy coding by reducing spatial correlations in the latent domain.

Inspired by this process, we consider the inverse formulation: given a random noise sample \( z \sim \mathcal{N}(0, I) \), we try to generate the latent code \( y \) which restores the spatial correlations. Our key idea is to generate the latent using a predicted Gaussian distribution, effectively reversing the normalization.

Following the reparameterization trick, the latent is reconstructed as:

\begin{equation}
y_{pred} = z \cdot \sigma_{pred} + \mu_{pred}, y_{pred} \sim \mathcal{N}(\mu_{pred}, \sigma_{pred}^2).
\end{equation}
To estimate the Gaussian parameters \( \mu_{\text{pred}} \) and \( \sigma_{\text{pred}} \), we introduce a lightweight Gaussian Parameter Prediction (GPP) module, which maps the input noise to these statistics in a spatially aware manner. The resulting latent \( y_{\text{pred}} \) is expected to capture essential image structures and texture information.

Given the difficulty of directly generating realistic latents from noise, we adopt an implicit neural representation (INR)-style training paradigm, where the model is overfitted to each individual input image. This setup removes the need for dataset-scale generalization and allows us to fully exploit the image-specific optimization, making the latent generation process both tractable and effective.
\begin{figure*}[t!]
    \centering
    \includegraphics[width=1\linewidth]{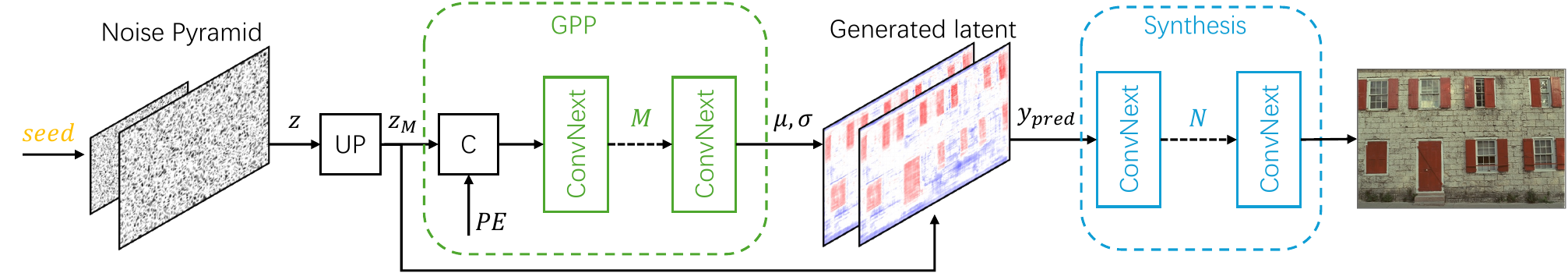}
    \caption{Overview of the proposed framework. $UP$ and $C$ denote upsampling and concatenation, respectively. Colored modules are included in the bitstream: the random seed, and the parameters of the GPP and synthesis networks.}
    \label{fig:diagram}
\end{figure*}

\section{Proposed Method}
\subsection{Overview}
The overall framework of the proposed method is illustrated in Fig.~\ref{fig:diagram}. Instead of explicitly storing or transmitting image-specific latent codes, we propose to generate the latent representation directly from a shared random noise tensor. Specifically, given a random seed, a multi-scale spatial noise pyramid \( z \) is deterministically generated at both the encoder and decoder sides. A lightweight module, referred to as the Gaussian Parameter Predictor (GPP), estimates the Gaussian parameters \( \mu_{\text{pred}} \) and \( \sigma_{\text{pred}} \) that define the latent distribution. These parameters are then used to synthesize the latent via the reparameterization trick, a technique commonly used in variational inference. The generated latent is subsequently fed into a synthesis network to reconstruct the image. Note that the random seed needs to be signaled to the decoder, enabling consistent noise generation without transmitting the latent itself.

\subsection{Noise Pyramid}
To provide rich spatial priors for latent generation, we construct a fixed multi-scale noise pyramid \( z = [z_1, z_2, \dots, z_N] \) based on a given random seed. Each \( z_i \) is a spatial tensor of shape \( [1, c_i, H/2^{i-1}, W/2^{i-1}] \), where \( c_i \) denotes the number of channels at scale \( i \), and \( H \), \( W \) are the height and width of the input image, respectively.

To form a unified representation, each \( z_i \) is upsampled to the resolution of the highest scale \( z_1 \) using bilinear interpolation. The upsampled tensors are then concatenated along the channel dimension, producing the final input noise tensor \( z_M \in \mathbb{R}^{1 \times C \times H \times W} \), where \( C = \sum_i c_i \). This noise tensor is fixed and deterministically generated from the seed, ensuring it remains unchanged throughout both training and inference.

\subsection{Gaussian Parameter Predictor}
To estimate the distribution of the latent representation from noise, we design a lightweight convolutional module termed the Gaussian Parameter Predictor (GPP). The input to the GPP is the concatenation of the multi-scale noise tensor \( z_M \in \mathbb{R}^{1 \times C \times H \times W} \) and a positional embedding tensor \( PE \in \mathbb{R}^{1 \times C_e \times H \times W} \) that encodes spatial information. This enriched input captures both structured randomness and spatial priors necessary for accurate modeling of image content:
\begin{equation}
\mu_{\text{pred}}, \sigma_{\text{pred}} = \mathrm{GPP}(\mathrm{concat}(z_M, PE))
\end{equation}

Here, \( \mu_{\text{pred}}, \sigma_{\text{pred}} \in \mathbb{R}^{1 \times C \times H \times W} \) represent the predicted mean and standard deviation maps for the latent distribution. The channel dimension \( C \) matches the dimensionality of the target latent space.

The GPP is implemented using a small number of ConvNeXt blocks~\cite{convnext}, balancing modeling capacity and computational efficiency.

Given the predicted Gaussian parameters, we generate the latent via the reparameterization trick:
\begin{equation}
y_{\text{pred}} = \mu_{\text{pred}} + \sigma_{\text{pred}} \cdot z_M
\end{equation}
This transformation enables one-shot latent generation from the shared noise input, eliminating the need to store or transmit latent codes.

\subsection{Synthesis Network}

The sampled latent $y_{pred}$ is fed into a synthesis network to reconstruct the target image $\hat{x}$:

\begin{equation}
\hat{x} = \mathrm{Synthesis}(y_{pred})
\end{equation}

The synthesis network is built on top of ConvNeXt blocks, which shows effective representation ability while maintaining low model complexity.

\begin{table}[t]  
\centering
\caption{Network configurations and complexity statistics of different settings.}
\label{tab:network_configs}
\small 
\begin{tabular}{lccccccc} 
\toprule
Settings & Scales & NCh & CCh & PE dims & $M$/$N$ & \#Params (k) & kMAC/pixel \\
\midrule
Setting 0 & 4 & 12 & 8  & 8  & 3/3 & 4.11 & 4.36 \\
Setting 1 & 4 & 12 & 10 & 10 & 3/3 & 6.11 & 5.81 \\
Setting 2 & 4 & 12 & 12 & 12 & 4/4 & 9.87 & 9.46 \\
Setting 3 & 4 & 12 & 16 & 10 & 3/3 & 12.58 & 12.17 \\
Setting 4 & 4 & 12 & 16 & 10 & 4/4 & 15.55 & 15.04 \\
\bottomrule
\end{tabular}
\vspace{6pt} 
\parbox{\linewidth}{\small \textit{Note:} NCh = noise channels; CCh= conv channels; PE dims = PE dimensions.}
\end{table}



\begin{figure}
    \centering
    \includegraphics[width=1.0\linewidth]{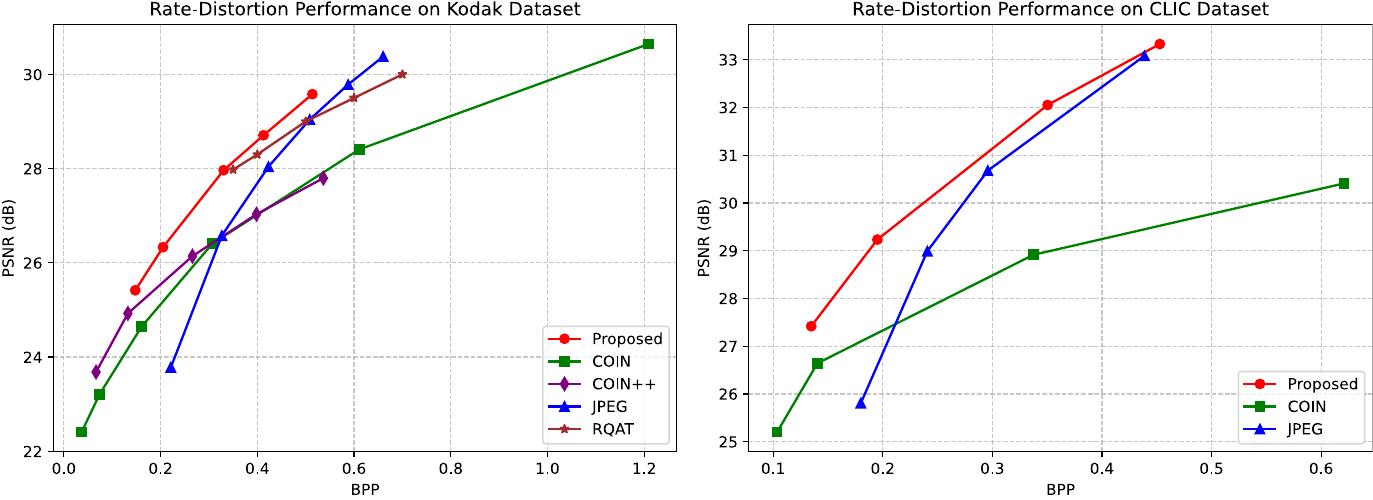}
    \caption{Rate-Distortion curves on Kodak and CLIC datasets.}
    \label{fig:rdcurve}
\end{figure}

\subsection{Network Compression}

We follow the Cool-Chic \cite{coolchic} to compress the network parameters of GPP and synthesis moduled. To achieve an optimal trade-off between rate and distortion, we perform a mesh search over a set of candidate quantization step sizes. For each candidate, the rate is estimated using Exponential-Golomb coding, while the distortion is measured in terms of PSNR between the reconstructed and original images. The quantization step that yields the best rate-distortion performance is selected for final encoding. Additionally, the random seed value is signaled as an integer value with 16 bits in the bitstream.

\subsection{Network Configurations and Complexity}

Table~\ref{tab:network_configs} summarizes the network architecture parameters used in our paper. All configurations employ 4 noise scales and a fixed channel count of 12 per scale. Variations include the convolutional channel width, positional embedding (PE) dimensionality, and the number of ConvNeXt blocks in the Gaussian Parameter Prediction (GPP) and synthesis modules, denoted by \(M\) and \(N\), respectively.

\section{Experiments}
\subsection{Experimental Setup}
\textbf{Training Details.} All models are trained using the Adam optimizer with an initial learning rate of $8 \times 10^{-3}$, and the learning rate is scheduled via cosine annealing. We define five complexity settings for the GPP and synthesis modules to explore different rate-distortion trade-offs. Following common practice in INR-based compression, all modules are trained in an overfitting manner on a single image, allowing the network to adapt its parameters to the specific image content without requiring large-scale datasets. To ensure stable training, we use Mean Squared Error (MSE) loss between the reconstructed image $\hat{x}$ and the input image $x$.

\textbf{Evaluation Method.} The proposed method is evaluated on the widly used datasets: Kodak \cite{kodak} and CLIC \cite{clic2020p} datasets. We report the peak signal-to-noise ratio (PSNR, in dB) and bitrate measured in bits per pixel (bpp). The bitrate accounts only for the compressed network parameters and the random seed, as no latent codes are transmitted in our framework.

\begin{figure}
    \centering
    \includegraphics[width=1.0\linewidth]{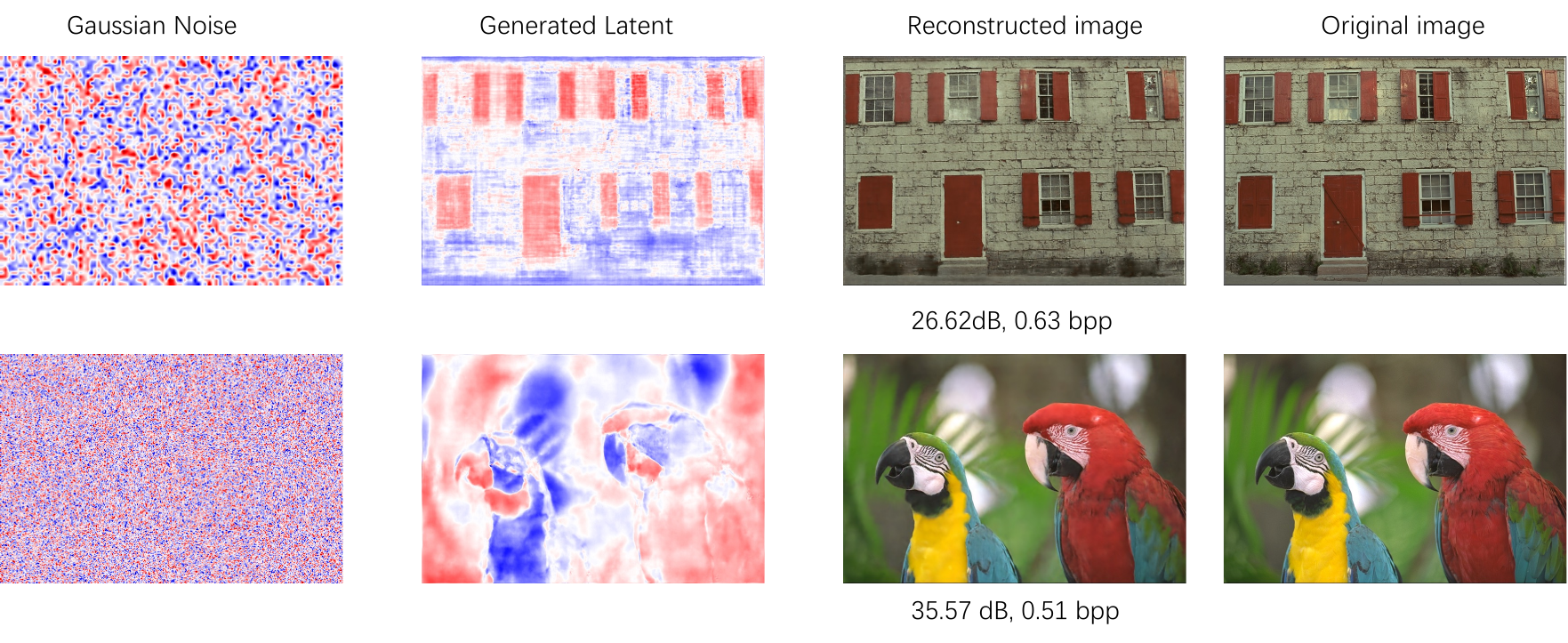}
    \caption{Visualization of key components in our framework. We extract one channel from the input noise and its corresponding channel from the generated latent for illustration. Best viewed in screen.}
    \label{fig:vis}
\end{figure}

\subsection{Rate-Distortion Performance}
We compare the proposed method with widely used image codec JPEG~\cite{jpeg} as well as several well-known INR-based image compression methods which use coordinates as input, including COIN~\cite{coin}, COIN++~\cite{coinpp}, RQAT-INR~\cite{RQAT}. The results are shown in \ref{fig:rdcurve}. As shown, our method consistently outperforms these baselines INR-based methods across a wide range of bitrates, demonstrating superior reconstruction quality and compression efficiency. Additionally, the proposed method achieves 25.74\% and 35.66\% BD-rate reduction on Kodak dataset when compared to JPEG and COIN, respectively.

\subsection{Visualization}
To gain deeper insight into our framework, we visualize intermediate representations at key stages: the input noise, the generated latent, and the reconstructed image. As shown in Fig.~\ref{fig:vis}, the Gaussian noise displays purely random structure without semantic content, as expected. After processing through the GPP module, the generated latent emerges with clear spatial patterns and structural cues that align closely with the target image. This highlights the model’s ability to capture image-specific priors from an initially structureless input. Finally, the synthesis network reconstructs a high-quality image from the generated latent, demonstrating that the learned transformation from noise to latent effectively encodes sufficient information for accurate image reconstruction.

These visualizations also validate our core hypothesis: given random Gaussian noise, when paired with a learned latent generation pipeline, can serve as an effective proxy for explicitly stored latent codes.

\begin{figure}
    \centering
    \includegraphics[width=0.85\linewidth]{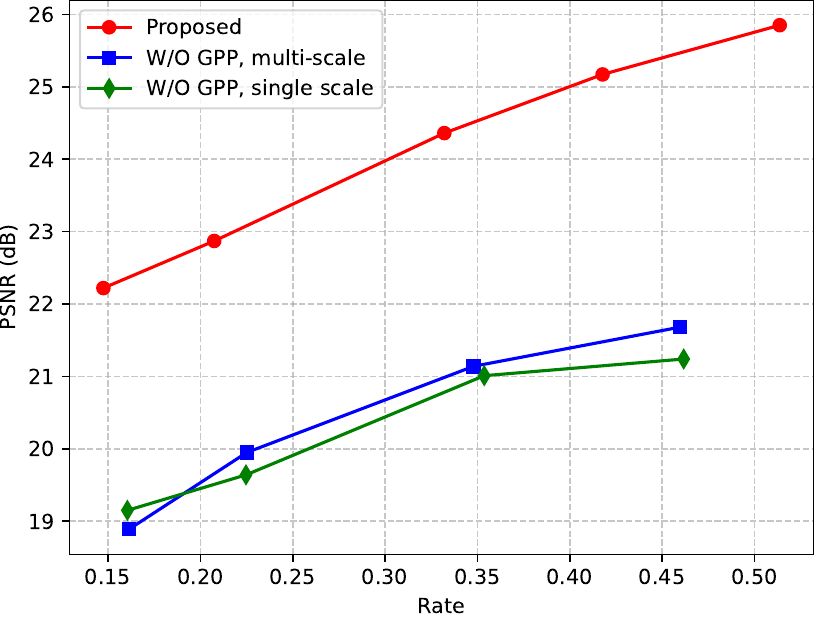}
    \caption{Rate-Distortion performance for ablation studies.}
    \label{fig:AB}
\end{figure}

\subsection{Ablation Studies}
To better understand the contribution of each design in our framework, we conduct several ablation studies in this section. 

\subsubsection{Effectiveness of GPP}
We first evaluate the importance of explicitly modeling the latent distribution via the GPP module. To this end, we compare our method with a baseline that directly maps noise to reconstructed images without generating the latent. Specifically, the GPP module is removed, and the synthesis network is slightly adjusted to achieve comparable bitrates. As shown in Fig.~\ref{fig:AB}, our method achieves significantly higher PSNR, demonstrating the effectiveness of Gaussian latent modeling in improving reconstruction quality.

\subsubsection{Noise Pyramid}
We analyze the impact of the input noise structure by comparing the proposed multi-scale noise pyramid against a single-scale noise input. Our results show that the multi-scale variant consistently outperforms the single-scale baseline in medium and high bitrates settings. This improvement is likely attributed to the richer spatial priors encoded by the hierarchical noise across low- and high-frequency components. Interestingly, under extremely low bitrates conditions, the single scale variant performs better, suggesting that the benefit of hierarchical noise may become more prominent as bitrates increases. \

\subsection{Comparison for Decoding Time}
We compare the proposed method (using Setting 4) with latent-storage approaches, including Cool-Chic, MLIC~\cite{mlic}, and Cheng2020~\cite{cheng2020}, in terms of decoding time. The results are presented in Table~\ref{tab:decoding_time}. As shown, our method achieves substantially faster decoding since it eliminates the need for the auto-regressive latent decoding required by Cool-Chic.

\begin{table}[t] 
    \centering
    \caption{Comparison of decoding times on Kodak dataset}  
    \label{tab:decoding_time}  
    \begin{tabular}{lcccc}  
        \toprule  
        \multicolumn{1}{c}{Metric}  & \multicolumn{1}{c}{Proposed} & \multicolumn{1}{c}{Cool-Chic v3.2}  & \multicolumn{1}{c}{MLIC}  & \multicolumn{1}{c}{Cheng2020} \\
        \midrule  
        CPU Time (ms) & 54 & 168 & 170 & 8658 \\
        \bottomrule 
    \end{tabular}
\end{table}

\subsection{Seed Robustness Analysis}
\begin{figure}
    \centering
    \includegraphics[width=0.85\linewidth]{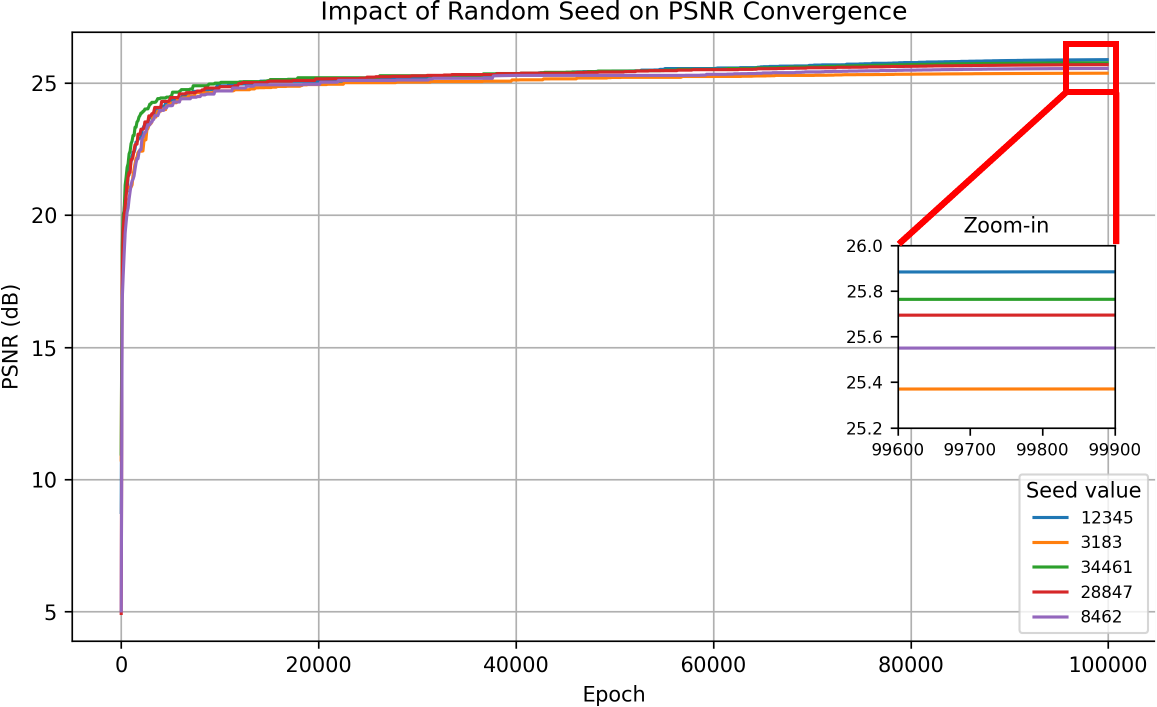}
    \caption{Influence of random seed values for training.}
    \label{fig:seed}
\end{figure}

We examine the impact of different random seeds used for noise generation in Fig.~\ref{fig:seed}. Specifically, we evaluate five distinct seed values on the $Kodim01.png$ image. While the convergence speed varies slightly among seeds, all models eventually stabilize to a narrow range of final PSNR values. This demonstrates that the proposed method is robust to the choice of seed. Notably, the performance gap between the best and worst final PSNR remains within 0.5 dB, suggesting that seed selection offers potential for further optimizing coding performance with minimal overhead.

\section{Conclusion}
In this paper, we proposed a novel INR-based image compression paradigm that generating the Gaussian latent from noise. Using a Gaussian parameter predictor and a synthesis network, the method achieves one-shot latent generation and image reconstruction via the reparameterization trick. Despite its simplicity, the approach yields competitive rate-distortion performance on Kodak and CLIC datasets, offering a promising trade-off between complexity and efficiency.


\Section{References}
\bibliographystyle{IEEEtran}
\bibliography{refs}

\end{document}